\documentclass[runningheads]{llncs}
\usepackage[utf8]{inputenc}
\usepackage[T1]{fontenc}
\usepackage{amsmath}

\usepackage{graphicx}
\usepackage[table]{xcolor}
\usepackage{tabularx}
\usepackage{booktabs}
\usepackage{multirow}
\usepackage{pmboxdraw}
\usepackage{amsmath}

\usepackage{inconsolata}  
\usepackage{pgfplotstable}
\usepackage{collcell}

\newcommand*{\MinNumber}{0.00}
\newcommand*{\MidNumber}{0.50}
\newcommand*{\MaxNumber}{1.00}

\definecolor{softred}{RGB}{255,140,140}
\definecolor{softyellow}{RGB}{255,255,150}
\definecolor{softgreen}{RGB}{140,255,140}

\newcommand{\ApplyGradient}[1]{%
        \ifdim #1 pt > \MidNumber pt
            \pgfmathsetmacro{\PercentColor}{max(min(100.0*(#1 - \MidNumber)/(\MaxNumber-\MidNumber),100.0),0.00)} %
            \hspace{-0.33em}\colorbox{softgreen!\PercentColor!softyellow}{#1}
        \else
            \pgfmathsetmacro{\PercentColor}{max(min(100.0*(\MidNumber - #1)/(\MidNumber-\MinNumber),100.0),0.00)} %
            \hspace{-0.33em}\colorbox{softred!\PercentColor!softyellow}{#1}
        \fi
}

\newcolumntype{G}{>{\collectcell\ApplyGradient}c<{\endcollectcell}}
\usepackage[table]{xcolor}
\usepackage{tabularx}
\usepackage{booktabs}
\usepackage{multirow}
\usepackage{pmboxdraw}
\usepackage{amsmath}
\begin{document}
\title{Hardware-Agnostic and Insightful Efficiency Metrics for Accelerated Systems: Definition and Implementation within TALP}
%
\titlerunning{Efficiency Metrics for Accelerated Applications}
%
\author{Ghazal Rahimi\inst{1}\orcidID{0009-0000-4946-5229} \and
Victor Lopez\inst{1}\orcidID{0000-0002-3113-9166} \and 
Marc Clasca\inst{1}\orcidID{0000-0002-5963-6658} \and 
Joan Vinyals Ylla Catal\`a\inst{1}\orcidID{0000-0002-4711-6815} \and 
Jesus Labarta\inst{1}\orcidID{0000-0002-7489-4727} \and 
Marta Garcia-Gasulla\inst{1}\orcidID{0000-0003-3682-9905}}

\authorrunning{G. Rahimi et al.}
%
\institute{Barcelona Supercomputing Center, Spain \\
\email{\{grahimi,vlopez,mclasca,jvinyals,jlabarta,martag\}@bsc.es}}
\maketitle              
\begin{abstract}
The increasing adoption of heterogeneous platforms that combine CPUs with accelerators such as GPUs in high-performance computing (HPC) introduces new challenges for performance analysis and optimization. Traditional efficiency metrics, such as those proposed by the Performance Optimization and Productivity (POP) Center of Excellence, were designed primarily for homogeneous CPU-based systems and therefore, do not capture the complex interactions between host and device resources. In this work, we extend the POP efficiency framework to heterogeneous architectures by introducing a new hierarchy of metrics that separately evaluate host and device efficiency. On the host side, we quantify the effectiveness of hybrid execution and offloading operations. On the device side, we propose a multiplicative hierarchy analogous to the host hierarchy and define its Parallel Efficiency branch. Beyond their definition and formulation, we present the implementation of these metrics in the TALP module of the DLB library. TALP is a lightweight monitoring library that provides measurements both post mortem and at runtime, with outputs available in textual and machine-readable formats. We validate the proposed framework through synthetic benchmarks and three production HPC applications, demonstrating how the metrics expose inefficiencies in offloading, load balance, and orchestration. Results show that the extended TALP metrics provide actionable insights to guide developers in optimizing heterogeneous HPC codes.


\end{abstract}
\section{Introduction}


From earth and life sciences to computer science and engineering, High Performance Computing (HPC) is rapidly transforming the world by helping scientists and engineers tackle complex challenges more efficiently.
Since 1965, Moore's law had continued to hold true, predicting that approximately every two years the number of transistors on integrated circuits would double, resulting in more efficient and higher-performing processors at little to no additional cost.

However, Moore's \cite{intel_mooreslaw} law is now reaching its limits due to physical and technological constraints such as dark silicon. To continue increasing parallelism and performance, new hardware in the form of accelerators, particularly Graphical Processing Units or GPUs, has been introduced. 

Heterogeneous computing, with architectures that combine CPUs and accelerators such as general purpose GPUs, has therefore become a suitable solution for meeting the high computational demands of both AI and HPC workloads. Systems equipped with GPUs can achieve extremely high floating-point operations per second (FLOPS) by utilizing thousands of small, simple cores that execute operations concurrently. The importance of GPUs in HPC is clearly visible by the most recent TOP500 list \cite{Top500}, where 9 out of the 10 most powerful ranked systems include an accelerated partition. This reflects a broader trend in HPC design, moving away from traditional general-purpose processors toward many-core accelerators with specialized memory and communication models.

With the advent of heterogeneous computing systems, the initial challenge for HPC application developers and users was to port their codes to these new architectures. More than 20 years after GPUs revolutionized the HPC landscape, a significant fraction of HPC codes have been successfully adapted to exploit them. The current challenge, however, lies in developing and maintaining portable code that can run efficiently across diverse heterogeneous and accelerator-based systems. 

For this reason, performance measurements for applications running on modern heterogeneous systems have become not only more challenging but also increasingly critical. The growing diversity of hardware devices, such as GPUs from different vendors adds significant complexity to the analysis process. Most available performance tools are vendor-specific and tailored to expose detailed information about the utilization of their respective hardware architectures. While this level of detail can be valuable, it often overwhelms users, making it difficult for non-experts to interpret results and identify the key issues or bottlenecks in their codes. The lack of standardized, portable, and user-friendly tools further complicates the task, leaving many developers without clear guidance on how to optimize performance in heterogeneous environments.

In this paper, we introduce a set of efficiency metrics for host and device architectures that extend the POP Parallel Efficiency framework. These new metrics are designed to be agnostic to both the underlying architecture and the programming model used to offload work to the device. They provide users with a comprehensive view of the primary sources of efficiency loss in their applications, covering both host and device execution. The metrics have been implemented within TALP~\cite{talp} (Tracking Application Live Performance), a lightweight profiling tool specifically developed to deliver parallel efficiency metrics during production-scale HPC runs.

The main contributions of this paper are the following:
\begin{itemize}
    \item \textbf{Extension of POP metrics to heterogeneous accelerated systems:} we complement the established POP methodology by introducing hardware-agnostic efficiency metrics that are suitable for accelerator-based architectures.

\item \textbf{Formal definition of the new metrics:} we provide a rigorous mathematical foundation to ensure their clarity, reproducibility, and applicability across platforms.

\item \textbf{Implementation within the TALP framework:} we deliver a proof-of-concept integration of the proposed metrics, currently applied to NVIDIA GPUs.

\item \textbf{Evaluation in different scenarios:} we validate the usefulness of the metrics through both a synthetic benchmark and three scientific production use cases.
\end{itemize}

The remaining of this paper is divided into the following sections, in Section \ref{sec:RW} we review the current state of the art of performance tools. In Section~\ref{sec:background} we introduce the original MPI POP metrics and the TALP framework. The new metrics, their formulation, interpretation and how they have been implemented in TALP are explained in Section~\ref{sec:metrics}. In Section~\ref{sec:evaluation}, we present the results of applying the metrics to different use cases. Finally, in Section~\ref{sec:conclusions} we will summarize the conclusions of the paper and the future work.

\section{Related Work}
\label{sec:RW}
Writing correct and functional codes that can scale and execute efficiently is becoming ever more complicated in High Performance Computing. Parallelism is the key requirement for running applications on all heterogeneous architectures, both in shared and distributed memory systems with nodes comprising of CPUs with or without attached GPUs. 
The European HPC community is driving rapid progress in computing, as evident from the TOP500 ranking, where five of the world's top 10 listed systems are located in Europe. One of the key features in this advancement is focusing on parallel computing performance and efficiency. Initiatives such as PRACE \cite{PRACE}(Partnership for Advanced Computing in Europe) have played a vital role in advancing HPC performance analysis by supporting tools and methodologies developed by the POP Center of Excellence (POP CoE)\cite{pop}. Over the course of three POP projects, they have helped users identify performance challenges and adopt best practices. A key element of this methodology is the use of performance analysis tools, which form the backbone of POP's approach to optimizing HPC applications. 

Each performance tool has its advantages and limitations, making them suitable for certain use cases and developers based on their features, needs or experience. A comprehensive overview of tools used in European HPC community is given in the white paper by Eriksson et al.\cite{whitepaper}, comparing all profiling and tracing tools. Here is a list of tools mentioned in this article:  1. \textit{HPCToolkit} \cite{HPCToolKit}  2. \textit{ITAC} \cite{ITAC}   3. \textit{Extrae and Paraver} \cite{extrae} \cite{paraver} and 4. \textit{SCALASCA} \cite{scalasca}. We briefly discuss the most relevant  tracing tools such as Extrae, Paraver and SCALASCA \cite{WYLIE2025107472}. Lastly, we introduce the most influential profiling tools such as TAU \cite{tau}, gprof \cite{gprof} and HPCToolkit \cite{HPCToolKit}. 

Paraver and Extrae \cite{extrae} are developed by Barcelona Supercomputing Center. Paraver is a data visualization browser for qualitative analysis and quantitative calculations. Extrae is a tracing tool, which uses several mechanisms to transparently capture performance data, ranging from dynamic instrumentation of unmodified production binaries to static linking. 
 
Scalasca is developed in Germany at the J\"ulich Supercomputing Center. Application codes for Scalasca \cite{scalasca} analysis are prepared via the Score-P \cite{scorep} instrumenter which configures adapters for various measurement interfaces and links associated measurement libraries.

The TAU \cite{tau} performance system is a profiling and tracing toolkit for performance analysis of hybrid parallel programs written in CUDA C, CUDA C\texttt{++}, OpenCL or using pyCUDA or OpenACC.  TAU gathers performance information of GPU computations and integrates it with other application performance data, through instrumentation of functions, methods, basic blocks, and statements to capture a performance picture of the resulting application execution.

Gprof \cite{gprof} is a performance analysis tool used to profile applications to determine where time is spent during program execution. Gprof is included with most Unix/Linux implementations, is simple to use, and can quickly show which parts of an application take the most time. Gprof works by automatically instrumenting your code during compilation, and then sampling the application's program counter during execution.

Lastly, we discuss HPCToolkit \cite{HPCToolKit}, an integrated suite of tools for performance measurement and analysis of applications on systems ranging from multicore desktops to GPU-accelerated supercomputers.
The toolkit offers insight into CPU and GPU executions and their possible inefficiencies by providing statistical sampling of timers and hardware counters on CPU and by monitoring activities and events on GPUs.

PRACE \cite{whitepaper} reports that tracing tools dominate the HPC ecosystem, with relatively few independent profiling tools available. However, performance analysis is reliant on both profiling and tracing tools as they serve distinct roles. A tracing tool records fine-grained and detailed timestamped events such as function calls, communication events, synchronization, and memory accesses over the whole execution. However, a profiling tool offers a high level statistical summary of the program's behavior and is specially useful in spotting load imbalance and inefficiencies both in CPUs and GPUs. In large-scale HPC applications, where full tracing can produce overwhelming amounts of data, profiling plays a critical role by first highlighting key performance hotspots. This allows users to focus tracing efforts more effectively, making the overall performance analysis process more scalable and manageable. HPCToolkit, TAU, and Score-P all support both profiling and tracing; in contrast, gprof only provides profiling capabilities. These tools measure durations of kernel executions, memory allocations/deallocations, memory copies, synchronization events, GPU time, and more. However, they do not inherently indicate whether measured times reflect normal behavior or performance bottlenecks, which can make their results difficult to interpret. In contrast, TALP not only reports raw execution times and metrics, enabling performance engineers to perform their own calculations and create custom metrics, but also provides meaningful, ready to use POP metrics, and their extension to GPU metrics, that are very helpful for analyzing HPC applications.

In addition, there are vendor-specific tools such as NVIDIA Nsight systems \cite{NVIDIANsightSystems}, a tracing tool, and Nsight Compute \cite{NVIDIANsightCompute}, a profiling tool. Although, these performance tools offer vast amount of information for applications running on CUDA-enabled devices, they are not suitable for other accelerators and vendors such as AMD or Intel. TALP and new TALP metrics offer support and assistance in performance analysis regardless of the vendor. In addition, the TALP profiler and the DLB library are open source, making it easier for any developer to use the metrics and apply changes, if necessary. 

\section{Background}
\label{sec:background}
\subsection{DLB}
\label{subsec:dlb}

DLB~\cite{dlb_web} (Dynamic Load Balancing) is a modular runtime library designed to enhance the performance and scalability of parallel applications in a non-intrusive and transparent manner. It provides seamless integration with widely used programming models such as MPI, OpenMP, and OmpSs, thereby ensuring portability across HPC environments. In addition, DLB exposes a user-level API that enables fine-grained control for advanced users and facilitates integration with external components such as resource managers.

The current version of DLB include three modules:

\begin{itemize}
 \item \textbf{LeWI} (Lend When Idle~\cite{dlb,mpi_x}): This module improves the load balance of MPI applications by using a second level of parallelism (OpenMP or OMPSs). It dynamically adjusts the number of threads employed by the shared-memory programming model, thereby optimizing the utilization of the node's computational resources. These adaptations are performed transparently to the application, requiring no modifications from the user.
 \item \textbf{DROM} (Dynamic Resource Ownership Management~\cite{drom}): Allows a resource manager, the user, or the application to change the distribution of threads among processes to maximize the performance or the efficiency during the execution without stopping the running processes.
 \item \textbf{TALP} (Tracking Application Live Performance): Provides lightweight, online monitoring of application performance with minimal overhead. By providing efficiency metrics in real time, TALP enables both users and tools to identify bottlenecks, guide optimization efforts, and support adaptive runtime decisions. The module is designed to be non-intrusive, portable, and complementary to traditional offline performance analysis methodologies.
\end{itemize}

\subsection{TALP}

TALP~\cite{talp} is a portable, extensible, lightweight, and scalable tool for parallel performance measurement and monitoring. It supports both online and post-mortem performance analysis: metrics can be collected during application execution to guide runtime decisions, or examined afterwards to evaluate the efficiency of full production runs.

TALP provides parallel efficiency metrics derived from the well-established POP methodology, complemented by hardware counter information and raw metrics that allow users to define and compute custom indicators.

The tool relies on standard mechanisms such as PMPI and OMPT, ensuring transparent use for applications without requiring code modifications. For more advanced scenarios, TALP also offers a user-level API that enables developers to annotate their code and obtain metrics for specific regions of interest.

The only requirement for deploying TALP is a system capable of preloading or overriding libraries at runtime. For instance, on Linux systems, the dynamic linker \verb|ld| supports preloading shared objects via the \verb|LD_PRELOAD| environment variable. In practice, collecting all performance metrics with TALP is as simple as dynamically loading the DLB library before the MPI symbols are resolved.

The post-mortem output of TALP is available both as plain text in a human-readable format and as a JSON file, enabling automated processing and integration with data analytics workflows.

\subsection{POP MPI Metrics}
\label{subsec:pop-metrics}

The main efficiency metrics collected by TALP are derived from the methodology proposed by the POP CoE project.

\begin{figure}[htbp]
\includegraphics[width=\textwidth]{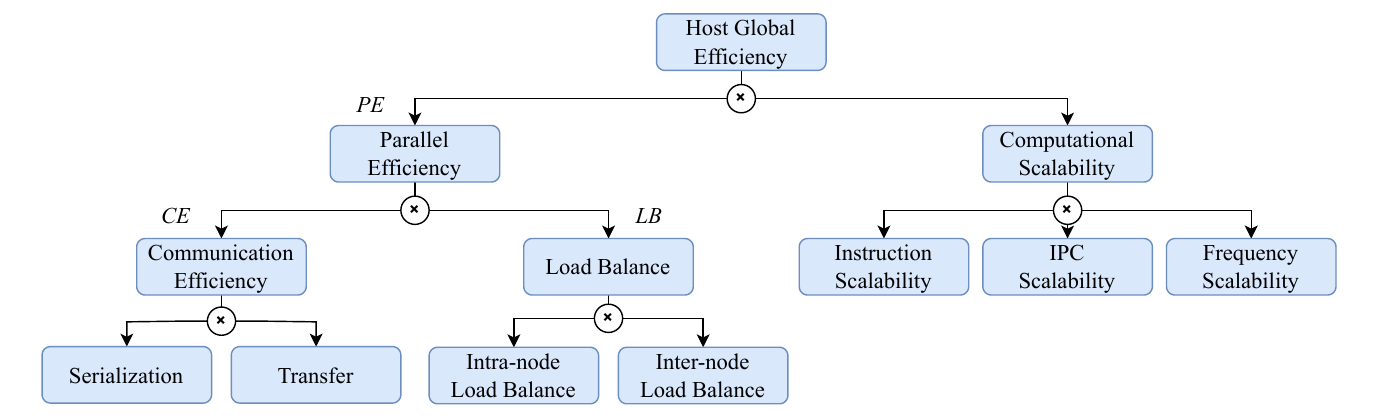}
\caption{Hierarchy of POP MPI metrics} \label{fig:pop_metrics}
\end{figure}

The POP metrics are organized hierarchically in a multiplicative structure, where each parent metric is defined as the product of its child metrics. Unlike traditional profiling approaches, they emphasize efficiency measurements that highlight performance losses rather than just reporting raw timing data. The hierarchical organization of the POP MPI metrics is shown in Figure~\ref{fig:pop_metrics}.

These metrics rely on a simplified execution model in which each process alternates between two states: a \textit{Useful} state, when it is performing computation, and a \textit{Not useful} state, when it is stalled (e.g., waiting in communication). In this section, as we focus on MPI applications, the \textit{Not useful} state corresponds specifically to the time spent in MPI calls.

The hierarchy distinguishes two main categories of metrics: Efficiency and Scalability. Efficiency metrics quantify the primary sources of performance loss within a given execution by measuring the fraction of \textit{Not useful} time. Scalability metrics, in contrast, are defined relative to a baseline case; they assess how the amount of \textit{Useful} computation evolves across different runs and attribute deviations to specific scaling factors.

Because TALP reports the metrics for a single run, only the efficiency metrics can be obtained. However, with the hardware counters collected by TALP, a user can compute the scalability metrics of several TALP runs.

In the remainder of this section, we define the efficiency metrics that will be extended in the following sections. All new metrics fall under the branch of Parallel Efficiency. However, the full metrics tree includes a second brach, Computational Scalability, which will not be discussed in this paper. 

We define $P = \{p_1, \dots, p_n\}$ as the set of MPI processes and $n$ the number of MPI processes. For each MPI process $p$ we define the set $U_p = \{u^p_1, u^p_2, \dots, u^p_{|U|}\}$ of the time intervals where the application is performing useful computation. We define $D_{U_p}$ as the sum of the durations of all useful time in a process $p$ (see equation (2)). Similarly, we can define $\overline{U}_p$ and $D_{\overline{U}_p}$ for not useful intervals. We also define the \textit{elapsed time} $E$ as shown in Equation~\ref{eqElapsed}.

\begin{equation}\label{eqElapsed}
E = max_{i=1}^{n}D_{U_i}+D_{\overline{U}_i}
\end{equation}

\begin{multicols}{2}
\noindent
\begin{equation}\label{eqUseful}
D_{U_p} = \sum_{j=1}^{|U_p|}u^p_j \quad
\end{equation}
\begin{equation}\label{eqPE}
PE = \frac{\sum_{i=1}^{n} D_{U_i}}{E * n}
\end{equation}
\end{multicols}

\begin{multicols}{2}
\noindent
  \begin{equation}\label{eqLB}
LB = \frac{\sum_{i=1}^{n} D_{U_i}}{n * \mbox{max}_{i=1}^{n} D_{U_i}}
\end{equation} \begin{equation}\label{eqCE}
CE = \frac{ \mbox{max}_{i=1}^{n} D_{U_i}}{E}
\end{equation}
\end{multicols}
\normalsize
 

The Parallel Efficiency metric ($PE$) indicates the amount of time lost due to parallelization, or equivalently, the ratio of useful computation time to total elapsed time, as shown in Equation~\ref{eqPE}. 

Parallel Efficiency has two child metrics: Load Balance ($LB$) and Communication Efficiency ($CE$), defined in Equations~\ref{eqLB} and~\ref{eqCE}.

Load Balance measures efficiency loss caused by uneven workloads (useful computation) across processes and is computed as the ratio between the average useful duration to the useful duration of the most loaded process.

Communication Efficiency quantifies the portion of time spent in communication that is not attributable to load imbalance.

\section{TALP Efficiency Metrics for Accelerated Platforms}
\label{sec:metrics}
\subsection{Metrics formulation and description}

\begin{figure}[htb!]
\centering
\includegraphics[width=\textwidth]{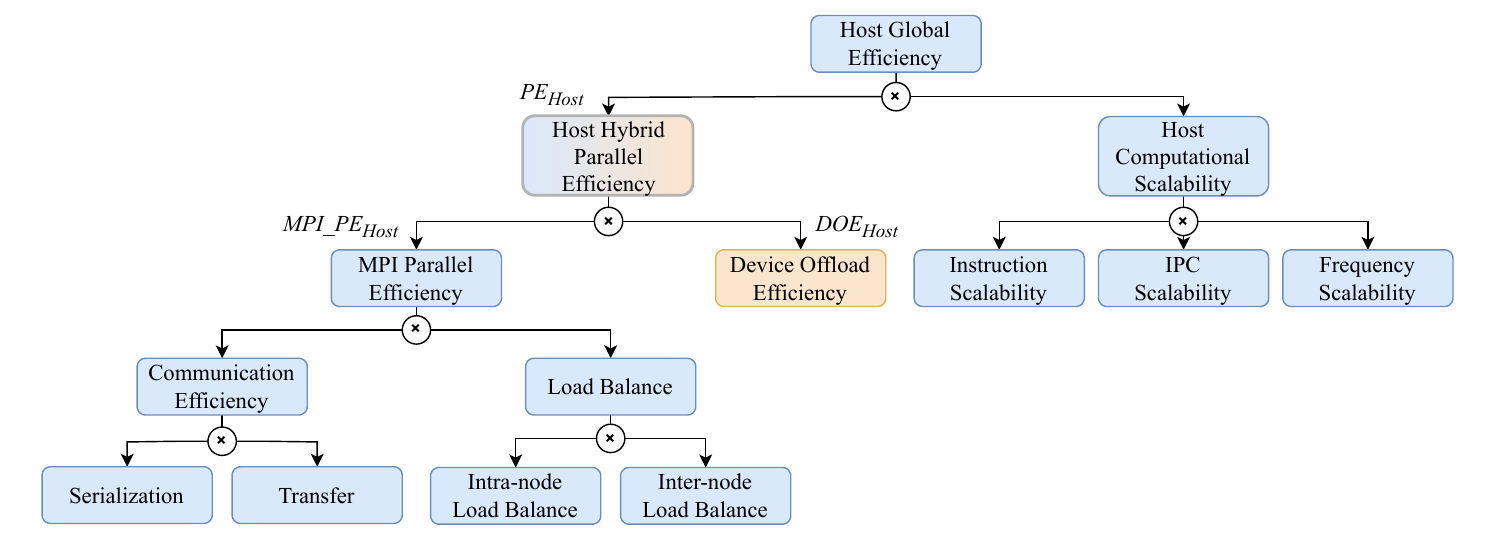}
\caption{Extended hierarchy of efficiency metrics for host resources. Orange boxes are the new added metrics} \label{fig:host_metrics}
\end{figure}

The efficiency metrics we propose for heterogeneous platforms extend the hierarchy introduced in Section~\ref{subsec:pop-metrics}. The key difference from the original formulation is that the hierarchy is divided into two disjoint trees, corresponding to host and device resources.

The first hierarchy, which extends the original POP model, evaluates the efficiency of host-side resource utilization. This structure is illustrated in Figure~\ref{fig:host_metrics}, where the newly introduced metrics are highlighted in orange: Host Hybrid Parallel Efficiency and Device Offload Efficiency. Host Hybrid Parallel Efficiency is defined as the product of MPI Parallel Efficiency and Offload Efficiency. The Offload Efficiency metric captures the fraction of efficiency lost when delegating work from the host to the device. This includes periods in which the CPU is blocked in GPU-related operations such as data transfers, kernel launches, and synchronization. Therefore, it quantifies inefficiencies arising from the use of accelerators.

\begin{figure}[htb!]
\centering
\includegraphics[width=0.7\textwidth]{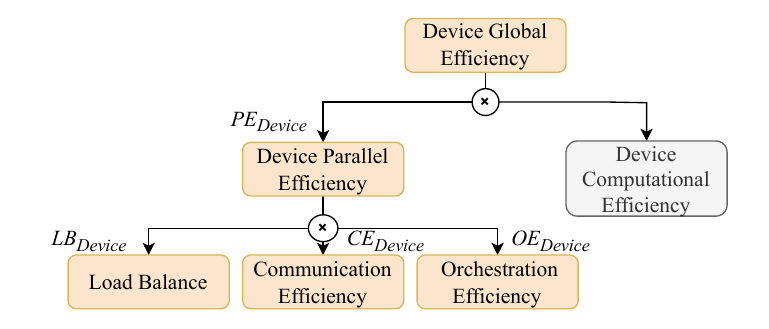}
\caption{Proposed hierarchy of efficiency metrics for device resources} \label{fig:device_metrics}
\end{figure}

The hierarchy of efficiency metrics for the device is presented in Figure~\ref{fig:device_metrics}. Analogous to the host-side hierarchy, device metrics are split into two branches: Device Parallel Efficiency, which measures whether device resources are being actively used, and Device Computational Efficiency, which measures how effectively those resources are exploited. In this paper, we focus on the Device Parallel Efficiency branch, leaving the definition of Device Computational Efficiency for future work.

Device Parallel Efficiency measures the fraction of time devices spend performing useful work, where useful work is defined as kernel execution. It is expressed as the product of three child metrics: Load Balance, Communication Efficiency, and Orchestration Efficiency.

Load Balance quantifies inefficiencies due to unequal kernel execution times across devices.
Communication Efficiency measures the efficiency loss caused by data transfers, whether between devices (GPU--GPU) or between host and device (CPU--GPU). Orchestration Efficiency captures idle periods where no useful work is scheduled, reflecting inefficiencies in coordinating computation, communication, and host offload.

As explained in Section~\ref{subsec:pop-metrics}, the original POP efficiency metrics are computed by classifying process execution into two states: \textit{Useful} and \textit{Not useful}. To extend these metrics to heterogeneous platforms, we define three states on the host and three states on the device.

For a process running on the host, the three states are: (i) \textit{Useful}, (ii) \textit{Device Offloading}, and (iii) \textit{MPI calls}. Let $D_{U_p}$ denote the total \textit{Useful} time of process $p$, and $D_{W_p}$ the total \textit{Device Offloading} time.

Host Hybrid Parallel Efficiency ($PE_{host}$) is defined as the ratio of useful computation time to the total elapsed CPU time (Equation~\ref{eqPE_host}).

\begin{equation}\label{eqPE_host}
PE_{host} = \frac{\sum_{i=1}^{n} D_{U_i}}{E \cdot n}
\end{equation}

MPI Parallel Efficiency ($MPI\_PE_{host}$) accounts for both useful and offloading times, treating them as \textit{Useful}, and divides their sum by the total elapsed CPU time (Equation~\ref{eqPE_mpi}); the formulas for its child metrics are omitted but apply the same treatment of states.

\begin{equation}\label{eqPE_mpi}
MPI\_PE_{host} = \frac{\sum_{i=1}^{n} (D_{U_i} + D_{W_i})}{E \cdot n}
\end{equation}

Device Offload Efficiency ($OE_{host}$) is computed as the ratio of useful computation time to the sum of useful and offloading times (Equation~\ref{eqOE}).

\begin{equation}\label{eqOE}
OE_{host} = \frac{\sum_{i=1}^{n} D_{U_i}}{\sum_{i=1}^{n} (D_{U_i} + D_{W_i})}
\end{equation}

For device metrics, we also classify execution into three states, treating each GPU as a single resource (streams are not distinguished here; this will be incorporated in future Device Computational Efficiency metrics). The three states are:
(i) \textit{Kernel computation} (useful work), (ii) \textit{Memory operations}, and (iii) \textit{Idle}. Importantly, since the metrics are defined at the device level, any overlap between computation and communication streams is counted as computation time.

Let $D_{K_g}$ be the time GPU $g$ spends executing kernels, $D_{M_g}$ the time spent on memory operations, and $m$ the number of devices. The device efficiency metrics are then defined as follows:

\begin{multicols}{2}
\begin{equation}\label{eqPEdevice}
PE_{device} = \frac{\sum_{i=1}^{m} D_{K_i}}{E \cdot m}
\end{equation}

\begin{equation}\label{eqLBdevice}
LB_{device} = \frac{\sum_{i=1}^{m} D_{K_i}}{m \cdot \max_{i=1}^{m} D_{K_i}}
\end{equation}
\end{multicols}

\begin{multicols}{2}
\begin{equation}\label{eqCEdevice}
CE_{device} = \frac{\max_{i=1}^{m} D_{K_i}}{\max_{i=1}^{m} (D_{K_i} + D_{M_i})}
\end{equation}

\begin{equation}\label{eqOEdevice}
OE_{device} = \frac{\max_{i=1}^{m} (D_{K_i} + D_{M_i})}{E}
\end{equation}
\end{multicols}

Device Parallel Efficiency is computed as the ratio between the useful time executing kernels and the total elapsed time (Equation~\ref{eqPEdevice}).

Load Balance is the ratio between the average useful duration across all devices and the maximum kernel duration (Equation~\ref{eqLBdevice}). 

Communication Efficiency is the ratio of the useful time on the most loaded device to the sum of its useful time and memory transfer time on the device with the least idle time (Equation~\ref{eqCEdevice}).

Orchestration Efficiency is the ratio of the process with the least idle time to the total elapsed time (Equation~\ref{eqOEdevice}).

\subsection{Implementation}

GPU efficiency metrics have been integrated into the TALP module of the DLB library through a modular, plugin-based design. The current implementation supports NVIDIA GPUs via CUDA and OpenACC, and AMD GPUs via HIP.

Two backends are currently implemented: one based on NVIDIA's CUPTI \cite{NVIDIACUPTI} interface (optionally complemented with OpenACC runtime hooks), and another based on AMD's rocprofilerv2 \cite{AMDRocProfiler} library. Each plugin is initialized during the startup of the main library and activates its components according to the available runtime environment. For instance, the OpenACC instrumentation is enabled only if the OpenACC library is detected.

Each plugin implements two complementary paths: (i) synchronous monitoring of host API calls through runtime callbacks, and (ii) asynchronous collection of device activity records through runtime-provided buffers. Host-side callbacks measure the duration of runtime calls such as kernel launches, memory transfers, or synchronization events, and immediately forward this information to the TALP module. Device-side records, on the other hand, are delivered in batches once buffers are filled or explicitly flushed, and require post-processing before classification into TALP metrics.

On NVIDIA platforms, host activity is monitored by enabling CUPTI's runtime callback domain (\verb|cuptiEnableDomain| and \verb|CUPTI_CB_DOMAIN_RUNTIME_API| argument), which triggers a callback on every CUDA runtime API call, such as kernel launches, memory operations, or synchronization calls. Device activity is collected through CUPTI's activity API (\verb|cuptiActivityEnable|) by enabling several activity kinds, including kernel execution, memory transfers, and memory operations.

When the OpenACC library is available, additional runtime events are intercepted through the \verb|acc_prof_register| interface, which registers callbacks for OpenACC-specific operations. This allows TALP to extend its visibility to applications that rely on directive-based programming.

On AMD platforms, we use \verb|rocprofiler_create_filter| to collect host activity with the \verb|ROCPROFILER_API_TRACE| parameter, combined with the call to the function \verb|rocprofiler_set_api_trace_sync_callback| to obtain callbacks for every HIP runtime API invocation. For device activity, a separate filter is created with the \verb|ROCPROFILER_DISPATCH_TIMESTAMPS_COLLECTION| 
parameter, which asynchronously delivers records of kernel executions.

Once device activity records are delivered, the plugin performs a uniform post-processing step, independent of the backend:
\begin{itemize}
\item Kernel execution records are flattened so that overlapping launches across streams are merged into a single continuous execution interval.
\item Memory transfer records are similarly flattened, and segments overlapping with kernel intervals are removed to avoid double counting.
\item Remaining uncovered regions are classified as inactive time, reflecting GPU underutilization.
\end{itemize}

The result is a precise breakdown of GPU state occupancy into computation, communication, and idle periods. This classification form the basis for the TALP device metrics, including Device Parallel Efficiency.

\section{Evaluation}

In this section, we demonstrate the practical applicability of the proposed metrics implemented in TALP by analysing a set of representative use cases. For each use case, we present the metrics generated by TALP, highlight the insights derived from the efficiency measurements, and compare them against a trace of the same execution to validate their accuracy.

We present two types of use cases. First, we employ a synthetic benchmark to construct simple execution patterns and report the corresponding metrics obtained. This allows us to validate the behavior of the metrics in controlled scenarios.

Second, we evaluate the extended version of TALP with three scientific production HPC codes. These real-world applications illustrate how the proposed metrics can reveal efficiency issues in large-scale, heterogeneous workloads.

For each use case, we provide both a table of the computed metrics and one or more execution traces. The traces serve as a visual confirmation that the reported metrics are consistent with the observed behavior, ensuring that the metrics are both relevant and representative of the workload characteristics.

\label{sec:evaluation}
\subsection{PILS}

PILS is a synthetic microbenchmark originally designed to emulate applications exhibiting various load imbalance scenarios. The initial implementation was developed in C and parallelized with MPI, OpenMP, and OmpSs. The updated version extends PILS for GPUs using CUDA while maintaining distributed execution with MPI. This extension enables the generation of controlled workloads involving both CPU and GPU computations, as well as memory operations.

In the experiments presented below, all use cases employ two MPI ranks, each bound to one CPU core and one GPU. Within the PILS benchmark traces, CPU states are visualized as follows: useful computation in blue, offloading tasks to the GPU in orange, and MPI communication in red. Device states are represented with useful kernel execution in blue, memory operations in green, and idle periods in gray.

\subsubsection{Use case 1: Loaded GPUs, underutilized CPUs, well balanced}
.\\

This use case represents an application with most of the work offloaded to devices, and CPUs only used to initialize, finalize and offload work to GPUs, the work is well balanced among the different GPUs and the different MPI processes running in the CPUs.

\begin{figure}[!htbp]
    \centering
    \includegraphics[width=0.7\columnwidth]{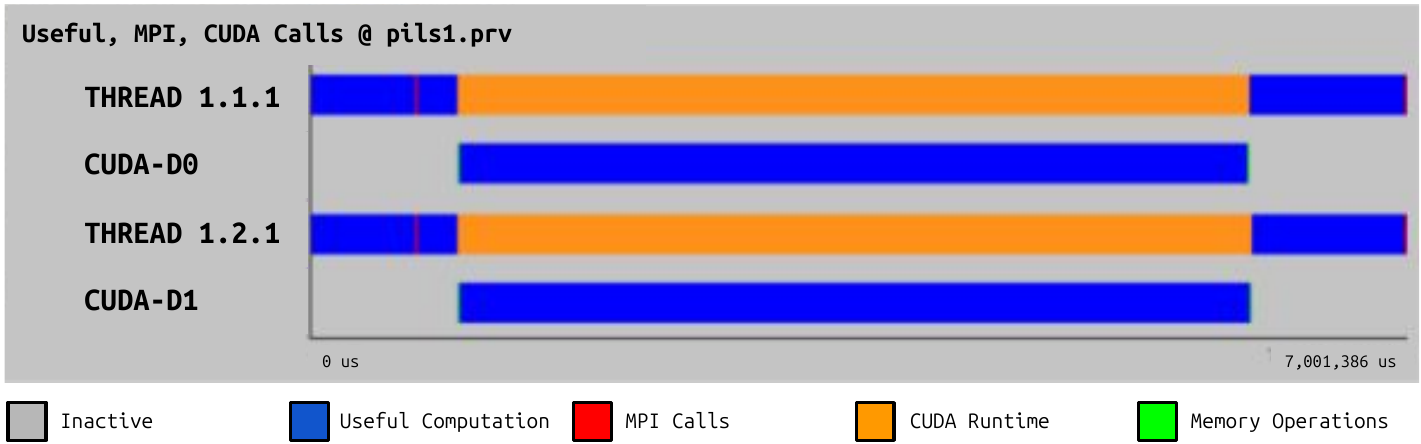}
        \begin{tabular}{p{.5cm}p{6cm}G}
        \toprule
        & \multicolumn{2}{l}{Metrics} \\
        \midrule
        \multirow{7}{*}{\rotatebox{90}{Host}}
        & \texttt{Parallel Efficiency}          & 0.16 \\
        & \texttt{├─ MPI Parallel Eff.}         & 1.00 \\
        & \texttt{│\ \ ├─ Comm. Eff.}           & 1.00 \\
        & \texttt{│\ \ └─ Load Balance}         & 1.00 \\
        & \texttt{└─ Device Offload Eff.}       & 0.16 \\
        \midrule

        \multirow{4}{*}{\rotatebox{90}{Device}}
        & \texttt{Parallel Efficiency}          & 0.82 \\
        & \texttt{├─ Load Balance}              & 1.00 \\
        & \texttt{├─ Communication Eff.}        & 1.00 \\
        & \texttt{└─ Orchestration Eff.}        & 0.82 \\
        \bottomrule
    \end{tabular}
    \caption{Use case 1: Loaded GPUs, underutilized CPUs, well balanced}
    \label{fig:PILS1}
\end{figure}

Figure~\ref{fig:PILS1} (top) shows the execution trace of this use case. We observe a nearly perfect balance between CPU and GPU workloads, within the respective hardware, and with GPU computation being approximately $10\times$ larger than CPU computation. The bottom part of Figure~\ref{fig:PILS1} presents the TALP output for this execution. All metrics report 100\% efficiency, with the exception of Device Offload Efficiency and Orchestration Efficiency.

From the host-side metrics, we conclude that the CPUs are primarily used to offload tasks to the GPUs rather than performing useful computation themselves, as reflected by the low Device Offload Efficiency. No significant MPI communication is observed in this case.

From the device-side metrics, we see that the GPUs are utilized efficiently. The Orchestration Efficiency of 82\% indicates that the GPUs experience some idle time while waiting for the CPUs to offload work.

A potential optimization for this pattern would be to overlap CPU computation with GPU kernel execution. This strategy would improve Device Offload Efficiency by making more effective use of host resources, while maintaining high GPU utilization.

\subsubsection{Use case 2: Loaded CPUs, underutilized GPUs, well balanced}
.\\

The second use case represents a scenario where the majority of the workload is executed on the host, with insufficient work offloaded to the device. The workload, however, remains well balanced across both CPUs and among GPUs.

\begin{figure}[!htbp]
    \centering
    \includegraphics[width=0.7\columnwidth]{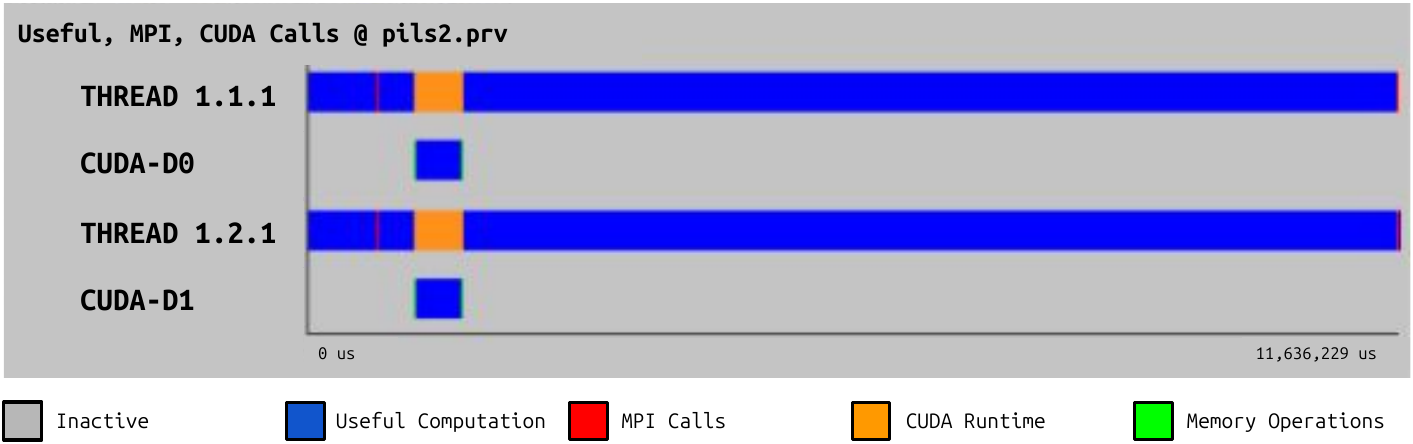}
        \begin{tabular}{p{.5cm}p{6cm}G}
        \toprule
        & \multicolumn{2}{l}{Metrics} \\
        \midrule
        \multirow{7}{*}{\rotatebox{90}{Host}}
        & \texttt{Parallel Efficiency}          & 0.94 \\
        & \texttt{├─ MPI Parallel Eff.}         & 1.00 \\
        & \texttt{│\ \ ├─ Comm. Eff.}           & 1.00 \\
        & \texttt{│\ \ └─ Load Balance}         & 1.00 \\
        & \texttt{└─ Device Offload Eff.}       & 0.94 \\
        \midrule

        \multirow{4}{*}{\rotatebox{90}{Device}}
        & \texttt{Parallel Efficiency}          & 0.05 \\
        & \texttt{├─ Load Balance}              & 1.00 \\
        & \texttt{├─ Communication Eff.}        & 1.00 \\
        & \texttt{└─ Orchestration Eff.}        & 0.05 \\
        \bottomrule
    \end{tabular}
    \caption{Use case 2: Loaded CPUs, underutilized GPUs, well balanced}
    \label{fig:PILS2}
\end{figure}

Figure~\ref{fig:PILS2} (top) shows the execution trace for this case. We observe that CPU computation is approximately $10\times$ greater than GPU computation, while both CPUs and GPUs are internally well balanced. The TALP output, shown in Figure~\ref{fig:PILS2} (bottom), reports excellent host-side efficiency metrics, all close to 100\%. The Device Offload Efficiency reaches 94\%, indicating that the CPUs spend most of their time performing useful computation, with only a small fraction devoted to offloading tasks or waiting for GPU kernels to complete.

In contrast, the device-side efficiency metrics are very low, with Device Parallel Efficiency at only 5\%. This reveals that GPUs are utilized for useful work only during a negligible portion of the execution time. 

From these results, we can conclude that the execution is dominated by host-side computation, while the accelerators remain largely underutilized. To improve performance, the application should be restructured to increase the fraction of the workload offloaded to GPUs, thereby leveraging their computational capacity.
\\
\\
\subsubsection{Use case 3: Loaded GPUs, imbalanced GPU computation, underutilized CPUs}
.\\

This use case models an application where most of the computation is offloaded to the GPUs, but the workload among devices is highly imbalanced.

\begin{figure}[!htbp]
    \centering
    \includegraphics[width=0.7\columnwidth]{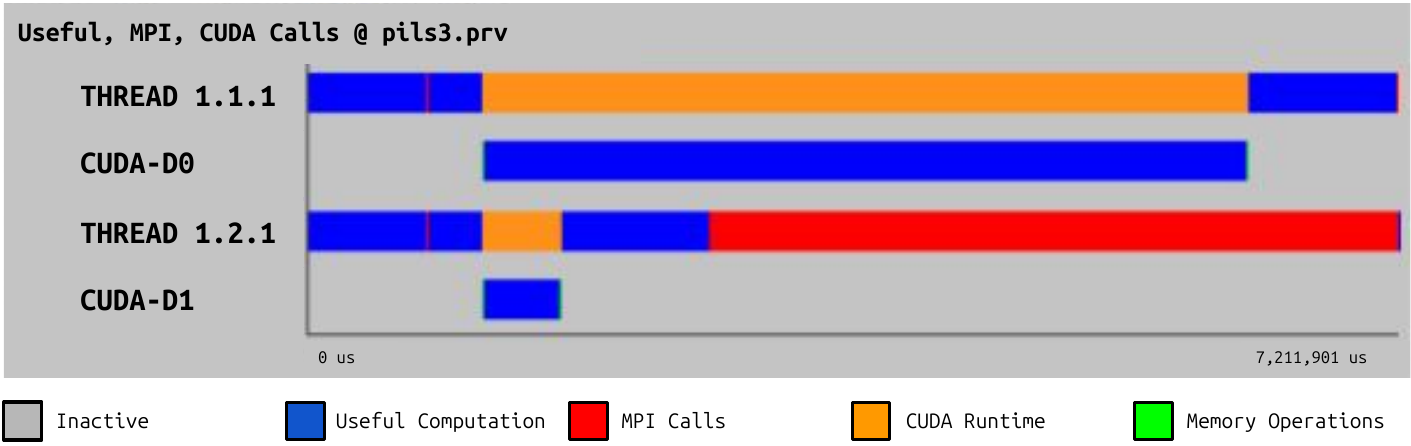}
        \begin{tabular}{p{.5cm}p{6cm}G}
        \toprule
        & \multicolumn{2}{l}{Metrics} \\
        \midrule
        \multirow{7}{*}{\rotatebox{90}{Host}}
        & \texttt{Parallel Efficiency}          & 0.16 \\
        & \texttt{├─ MPI Parallel Eff.}         & 0.63 \\
        & \texttt{│\ \ ├─ Comm. Eff.}           & 1.00 \\
        & \texttt{│\ \ └─ Load Balance}         & 0.63 \\
        & \texttt{└─ Device Offload Eff.}       & 0.26 \\
        \midrule

        \multirow{4}{*}{\rotatebox{90}{Device}}
        & \texttt{Parallel Efficiency}          & 0.45 \\
        & \texttt{├─ Load Balance}              & 0.55 \\
        & \texttt{├─ Communication Eff.}        & 1.00 \\
        & \texttt{└─ Orchestration Eff.}        & 0.82 \\
        \bottomrule
    \end{tabular}
    \caption{Use case 3: Loaded GPUs, imbalanced GPU computation, underutilized CPUs}
    \label{fig:PILS3}
\end{figure}

In the trace shown at the top of Figure~\ref{fig:PILS3}, we observe that GPU0 (from MPI rank 0) executes approximately $10\times$ more work than GPU1 and both CPUs. As highlighted in orange, MPI rank 0 spends most of its time in CUDA-related calls, whereas MPI rank 1 (running on CPU1) spends the majority of its time in MPI communication (red), waiting for rank 0 to complete.

The TALP metrics shown at the bottom of Figure~\ref{fig:PILS3} report very low overall efficiency. On the host side, the Device Offload Efficiency is only 26\%, indicating that CPUs spend most of their time idle, waiting for the GPUs to complete their workloads rather than performing useful work. Similarly, the Host MPI Parallel Efficiency is low, reflecting the imbalance between the two ranks. It is interesting to notice, that although there is no load imbalance between the useful work computed by CPUs it still shows load imbalance at the MPI level, this is due to the fact that we take into account the time used to offload work to GPUs to compute the load imbalance at the host side. This is intended to be like this because we assume that the work offloaded to the GPUs are work assigned to that MPI rank, therefore, the load assigned to the different MPI ranks is unbalanced as indicated by the efficiency metrics.

On the device side, the Load Balance metric is 55\%, confirming that GPU workloads are unevenly distributed: while one GPU is fully loaded, the other remains mostly idle for large portions of the execution. 

Improving the workload distribution across devices would substantially increase both Device Load Balance and Host MPI Parallel Efficiency, while better orchestration of computation and data transfers would help reduce idle time on the GPUs.

\subsubsection{Use case 4: Imbalanced GPUs and CPUs, CPUs more loaded than GPUs}
.\\

Use case 4 represents an application with load imbalance both at the host and device level.

\begin{figure}[!htbp]
    \centering
    \includegraphics[width=0.7\columnwidth]{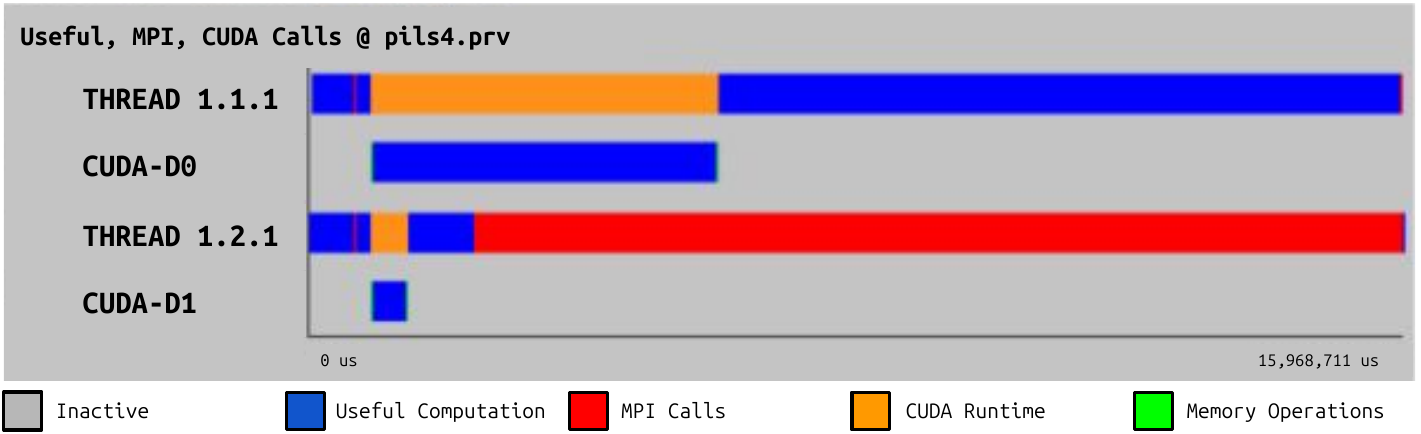}
        \begin{tabular}{p{.5cm}p{6cm}G}
        \toprule
        & \multicolumn{2}{l}{Metrics} \\
        \midrule
        \multirow{7}{*}{\rotatebox{90}{Host}}
        & \texttt{Parallel Efficiency}                 & 0.36 \\
        & \texttt{├─ MPI Parallel Eff.}          & 0.55 \\
        & \texttt{│\ \ ├─ Comm. Eff.}    & 1.00 \\
        & \texttt{│\ \ └─ Load Balance}                & 0.55 \\
        & \texttt{└─ Device Offload Eff.}        & 0.65 \\
        \midrule

        \multirow{4}{*}{\rotatebox{90}{Device}}
        & \texttt{Parallel Efficiency}                 & 0.18 \\
        & \texttt{├─ Load Balance}                     & 0.55 \\
        & \texttt{├─ Communication Eff.}         & 1.00 \\
        & \texttt{└─ Orchestration Eff.}         & 0.33 \\
        \bottomrule
    \end{tabular}
    \caption{Use case 4: Imbalanced GPUs and CPUs, CPUs more loaded than GPUs}
    \label{fig:PILS4}
\end{figure}

In the trace shown in Figure~\ref{fig:PILS4} top, we can see that the MPI rank 0 offloads a long chunk of work to its GPU (blue), waits for its completion (orange) and then has a long computation (blue). While MPI rank 1, has much less work offloaded to the GPU, after waiting for its GPU, MPI rank 1 has a short burst of compute time (blue), and waits for MPI rank 0 afterwards (red).

The output of TALP shown at the bottom of Figure~\ref{fig:PILS4} indicates a low Parallel Efficiency on the host. The Load Balance is 55\% showing that there is a very high load imbalance between the two MPI ranks. The Device Offload Efficiency indicates that CPUs are being underutilized while waiting for GPU computation to finish.
On the other hand, the device metrics show a Load Balance on 55\% indicating that the workload of the devices is not well distributed. The low Orchestration Efficiency indicates that the GPUs are underutilized because not enough work have been offloaded to them, as can be seen by the long chunk of work done by MPI rank 0 at the end of the execution.

In this case it would be crucial to distribute better the workload among CPUs and GPUs to improve the performance of the code. 

\subsubsection{Use case 5: Imbalanced CPU load, same global load CPU and GPU}
.\\

\begin{figure}[!htbp]
    \centering
    \includegraphics[width=0.7\columnwidth]{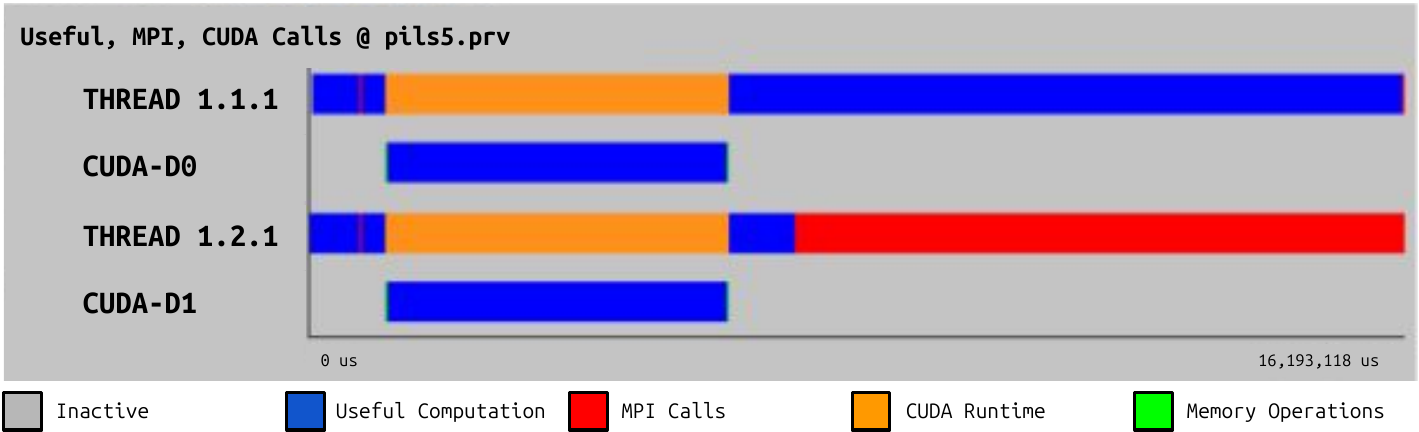}
        \begin{tabular}{p{.5cm}p{6cm}G}
        \toprule
        & \multicolumn{2}{l}{Metrics} \\
        \midrule
        \multirow{7}{*}{\rotatebox{90}{Host}}
        & \texttt{Parallel Efficiency}                 & 0.36 \\
        & \texttt{├─ MPI Parallel Eff.}          & 0.70 \\
        & \texttt{│\ \ ├─ Comm. Eff.}    & 1.00 \\
        & \texttt{│\ \ └─ Load Balance}                & 0.70 \\
        & \texttt{└─ Device Offload Eff.}        & 0.51 \\
        \midrule

        \multirow{4}{*}{\rotatebox{90}{Device}}
        & \texttt{Parallel Efficiency}                 & 0.33 \\
        & \texttt{├─ Load Balance}                     & 1.00 \\
        & \texttt{├─ Communication Eff.}         & 1.00 \\
        & \texttt{└─ Orchestration Eff.}         & 0.33 \\
        \bottomrule
    \end{tabular}
    \caption{Use case 5: Imbalanced CPU load, same global load CPU and GPU}
    \label{fig:PILS5}
\end{figure}

This pattern represents an application where the workload of the CPUs and the GPUs is the same, but the CPU load is not evenly distributed among the two MPI ranks. We can observe the pattern in the trace in Figure~\ref{fig:PILS5}. It can be seen that both MPI ranks offload work to their corresponding GPUs, and wait for them to finish it (orange). After the GPU computation is finished, there is an imbalanced chunk of work among the two MPI ranks, where MPI rank 1 waits for MPI rank 0 to finish (red).

The TALP metrics reported at the bottom of Figure~\ref{fig:PILS5} indicate a low Parallel Efficiency both for the host and the device.
In the case of the host, we see both a low Load Balance (70\%), pointing to the uneven workload distribution on the two MPI ranks, and a low Device Offload Efficiency, indicating that CPUs are not used to do useful computation while waiting for the GPUs to finish their work.

The device efficiency metrics show an Orchestration Efficiency of 33\%, indicating that not enough work is offloaded to the devices, and they are most of the time idle waiting for work.

This pattern could be improved by distributing better the workload of the MPI ranks and also by offloading more work from the CPUs to the GPUs.

\subsubsection{Use case 6: Even distribution of work, large host-device data movement}
.\\

Use case 6 shows a pattern were the computation is well distributed among GPUs and CPUS, most of the work is offloaded to the GPUs, but there is a large data movement between host and device.

\begin{figure}[!htbp]
    \centering
    \includegraphics[width=0.7\columnwidth]{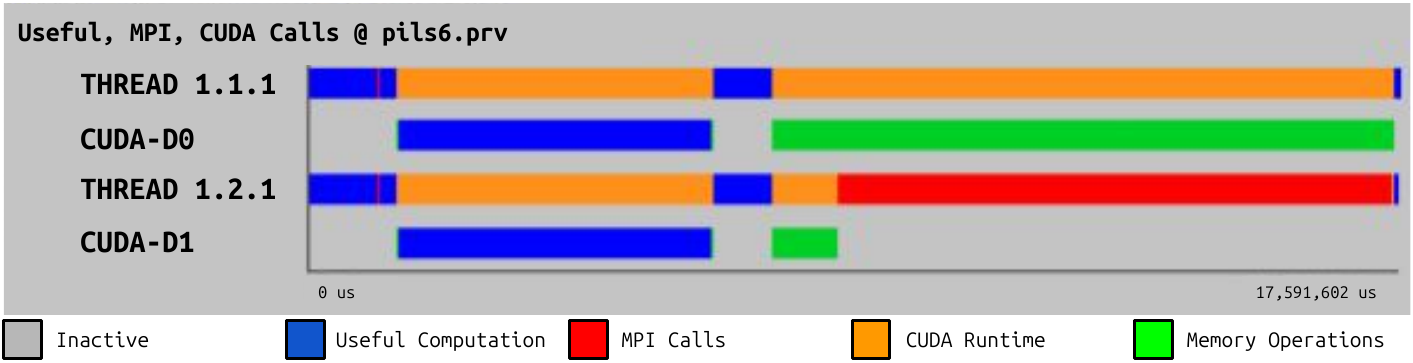}
        \begin{tabular}{p{.5cm}p{6cm}G}
        \toprule
        & \multicolumn{2}{l}{Metrics} \\
        \midrule
        \multirow{7}{*}{\rotatebox{90}{Host}}
        & \texttt{Parallel Efficiency}                 & 0.06 \\
        & \texttt{├─ MPI Parallel Eff.}          & 0.72 \\
        & \texttt{│\ \ ├─ Comm. Eff.}    & 1.00 \\
        & \texttt{│\ \ └─ Load Balance}                & 0.72 \\
        & \texttt{└─ Device Offload Eff.}        & 0.09 \\
        \midrule

        \multirow{4}{*}{\rotatebox{90}{Device}}
        & \texttt{Parallel Efficiency}                 & 0.31 \\
        & \texttt{├─ Load Balance}                     & 1.00 \\
        & \texttt{├─ Communication Eff.}         & 0.36 \\
        & \texttt{└─ Orchestration Eff.}         & 0.86 \\
        \bottomrule
    \end{tabular}
    \caption{Use case 6: Even distribution of work, large host-device data movement}
    \label{fig:PILS6}
\end{figure}

In Figure~\ref{fig:PILS6} (top) we can see the trace representing this use case. We can see how both MPI ranks show the same behavior until the end of the execution where MPI rank 0 needs to move a large chunk of data from the device (green), while MPI rank 1 waits in an MPI blocking call (red).

The TALP metrics obtained (Figure~\ref{fig:PILS6} bottom) show a very low Parallel Efficiency on the host. The main bottleneck is the Device Offload Efficiency showing an efficiency of 9\%, this indicates that the CPU is most of the time waiting for the GPUs. The host Load Balance is 72\%, indicating that the two MPI ranks do not have the same amount of work, in this case it is pointing at the data movement that is done only by one of the MPI ranks.

The device metrics show a Communication Efficiency of 36\%, indicating that there is an important efficiency loss due to data movements. The Orchestration Efficiency of 86\% shows that there is still room for improvement to offload work to the GPUs more efficiently by the host.

\subsubsection{Use case 7: Comparison of CPU-GPU computation overlap}
.\\

In this use case, we demonstrate how overlapping CPU and GPU workloads are shown in the newly introduced POP metrics. 

\begin{figure}[!htb]
    \centering
    No Overlap CPU-GPU computation\\
    \includegraphics[width=0.7\columnwidth]{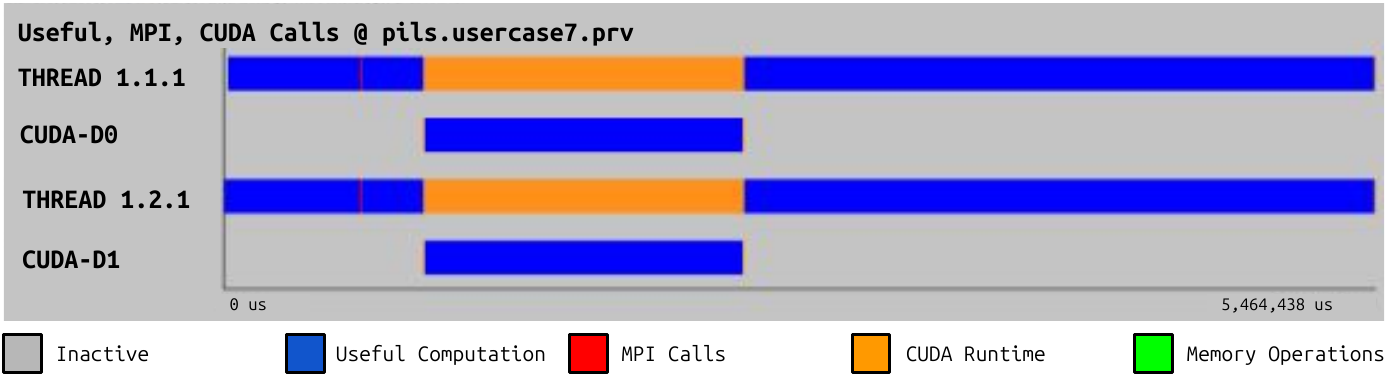}

    Overlap CPU-GPU computation
    \includegraphics[width=0.7\columnwidth]{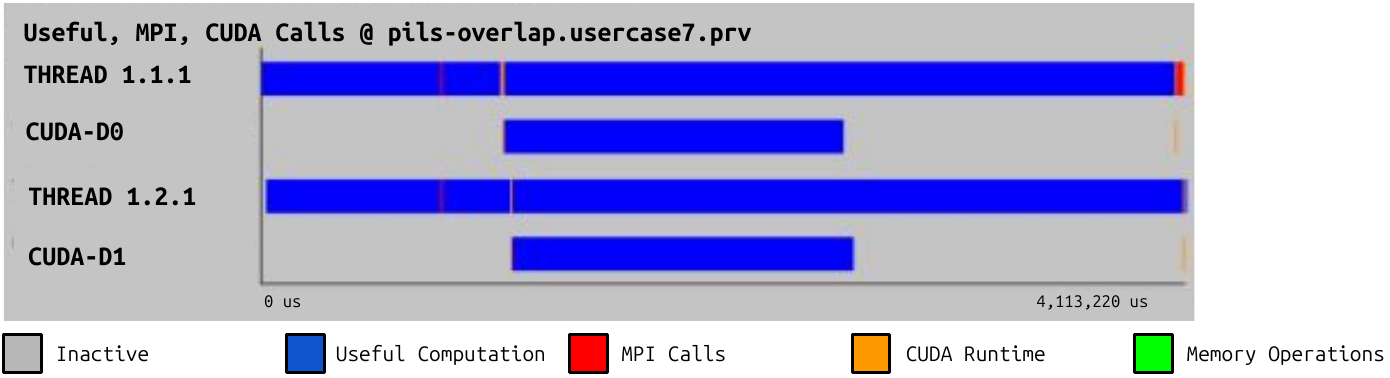}
    
        \begin{tabular}{p{.5cm}p{6cm}GG}
        \toprule
        & \multicolumn{1}{l}{Metrics} & \multicolumn{1}{c}{Seq} & \multicolumn{1}{c}{Overlap}\\
        \midrule
        \multirow{7}{*}{\rotatebox{90}{Host}}
        & \texttt{Parallel Efficiency}        &  0.64 &  0.97 \\
        & \texttt{├─ MPI Parallel Eff.}       &  1.00 &  1.00 \\
        & \texttt{│\ \ ├─ Comm. Eff.}         &  1.00 &  1.00 \\
        & \texttt{│\ \ └─ Load Balance}       &  1.00 &  1.00 \\
        & \texttt{└─ Device Offload Eff.}     &  0.64 &  0.97 \\
        \midrule
        \multirow{4}{*}{\rotatebox{90}{Device}}
        & \texttt{Parallel Efficiency}        &  0.33 &  0.49 \\
        & \texttt{├─ Load Balance}            &  1.00 &  1.00 \\
        & \texttt{├─ Communication Eff.}      &  1.00 &  1.00 \\
        & \texttt{└─ Orchestration Eff.}      &  0.33 &  0.49 \\
        \bottomrule
    \end{tabular}
    \caption{Use case 7: Comparison of CPU-GPU computation overlap.}
    \label{fig:PILS7}
\end{figure}

In Figure~\ref{fig:PILS7} top and middle we can see the two traces representing this use case, at the top there is no overlap between CPU and GPU computation, while in the middle trace the execution overlaps the CPU and GPU computations. 
In both traces, both CPUs and GPUs exhibit load balance across their respective resources. 

Looking at Figure~\ref{fig:PILS7} bottom we can see the TALP metrics obtained for both executions. We observe that the only metrics that vary between the two executions are Device Offload Efficiency and Orchestration Efficiency, which influence Parallel Efficiency on the host and device, respectively. Device Offload Efficiency improves by 33\%, reaching near-optimal levels in the overlapping execution. This improvement occurs because CPUs offload work to GPUs asynchronously for a short period, allowing them to spend the remainder of the time in useful computation. Orchestration Efficiency, on the other hand, reaches nearly 50\% in the overlapping execution, since the CPU workload is twice as large as the GPU workload, leaving the GPU inactive for half of the execution time.

\subsection{Scientific HPC applications}

In this section, we use three HPC scientific applications to test and validate the proposed metrics and their implementation. We provide a brief introduction to each application in the following subsections.

The applications are executed on the accelerated partition of MareNostrum5 (MN5-Acc)~\cite{MN5}, each node of MN5-Acc is equipped with $2\times$ Intel Sapphire Rapids 8460Y CPUs and $4\times$ NVIDIA H100 GPUs.

We profile all of them with TALP to identify performance bottlenecks, and the results are validated with Paraver traces.

\subsubsection{SOD2D}
.\\

 SOD2D implements a numerical solution for the governing equations of compressible and incompressible fluid flow in three dimensions. It is based on the spectral element method (SEM) and is intended for scale-resolving simulations, including large eddy simulation (LES) and direct numerical simulation (DNS). The implementation is written in Fortran and employs MPI to distribute work among the different nodes and cores and OpenACC to offload work to GPUs. 

\begin{table}[!htbp]
    \centering
    \caption{TALP Output for SOD2D from 1 to 8 nodes}
    \label{tab:sod2d}
    \begin{tabular}{p{.5cm}p{4cm}GGGG}
        \toprule
        & \multicolumn{1}{l}{Metrics} & \multicolumn{4}{c}{ Nodes} \\
        \cmidrule(lr{0.5em}){3-6}
        & 
        & \multicolumn{1}{c}{1}
        & \multicolumn{1}{c}{2}
        & \multicolumn{1}{c}{4}
        & \multicolumn{1}{c}{8}    \\
        
        \midrule
         \multirow{7}{*}{\rotatebox{90}{Host}}
         & \texttt{Parallel Efficiency}    & 0.06 & 0.05 & 0.04 & 0.04 \\
         & \texttt{├─ MPI Parallel Eff.}   & 0.94 & 0.88 & 0.79 & 0.67 \\
         & \texttt{│\ \ ├─ Comm. Eff.}     & 0.95 & 0.89 & 0.80 & 0.68 \\
         & \texttt{│\ \ └─ Load Balance}   & 1.00 & 0.98 & 0.99 & 0.99 \\
         & \texttt{└─ Device Offload Eff.} & 0.06 & 0.05 & 0.06 & 0.06 \\
         \midrule

         \multirow{4}{*}{\rotatebox{90}{Device}}
         & \texttt{Parallel Efficiency}    & 0.87 & 0.81 & 0.72 & 0.59 \\
        & \texttt{├─ Load Balance}         & 1.00 & 0.98 & 0.99 & 0.99 \\
        & \texttt{├─ Communication Eff.} & 0.99 & 0.99 & 0.99 & 0.99 \\
        & \texttt{└─ Orchestration Eff.} & 0.88 & 0.84 & 0.73 & 0.60 \\
        \bottomrule
    \end{tabular}
\end{table}

In Table~\ref{tab:sod2d} we show the efficiency metrics obtained with TALP when running SOD2D from 1 to 8 nodes and 1001 timesteps. We can see that the application is optimized for execution on GPUs, as reflected by the high Device Parallel Efficiency, while the host Device Offload Efficiency is extremely low. This indicates that GPUs remain active and engaged in useful work up to eight nodes while the CPUs are only used to offload work to the GPU and remain idle most of the time.

At 8 nodes, Orchestration Efficiency exhibits a significant drop. This behavior can likely be attributed to CPUs spending more time in inter-CPU communication, as evidenced by the decrease in host-side Communication Efficiency, thereby delaying the delegation of work to GPUs.

\begin{figure}[!htbp]
    \centering
    \includegraphics[width=0.9\columnwidth]{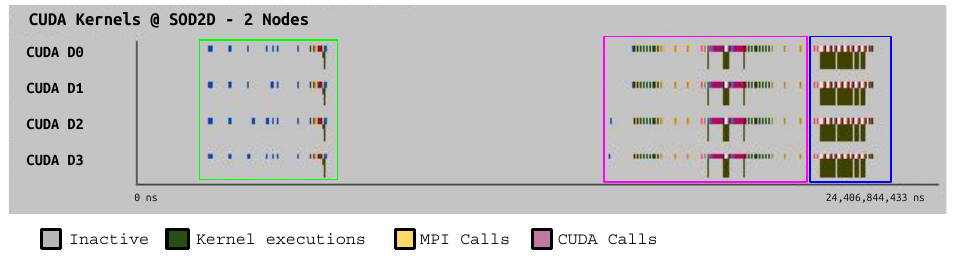}
    \includegraphics[width=0.9\columnwidth]{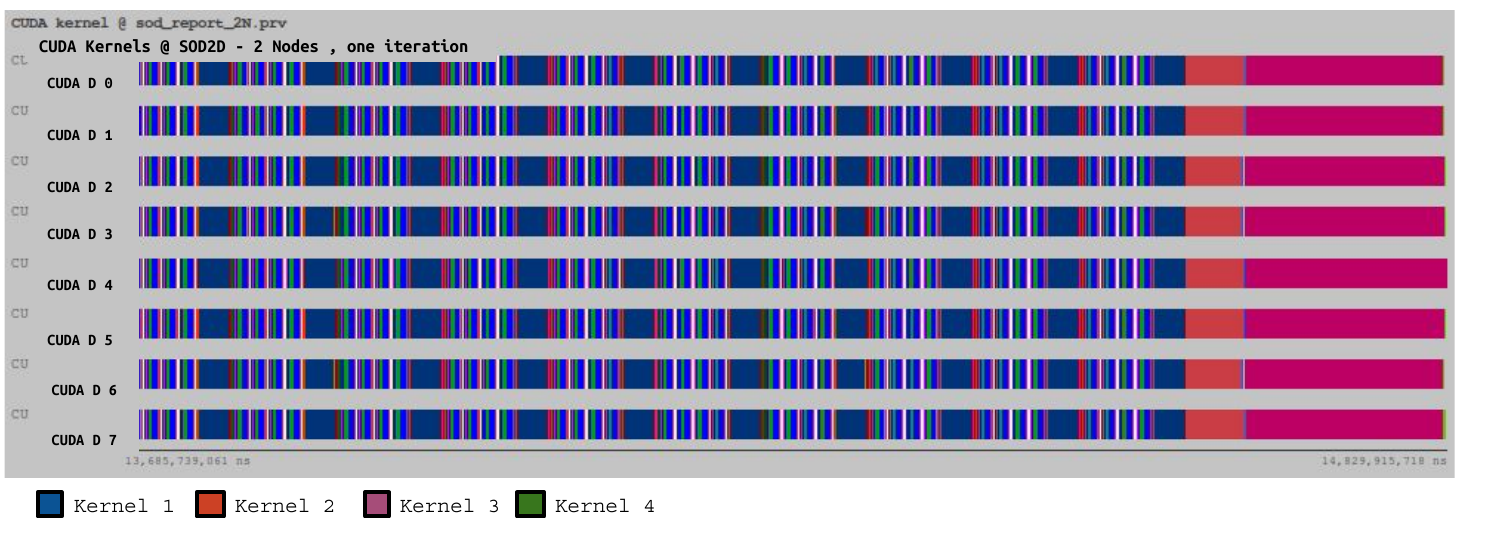}
    \caption{SOD2D trace: CUDA Kernel Activity. Top: full trace (11 timesteps) using 2 nodes and 8 GPUs. Bottom: detailed view of a single timestep from the same execution.}
    \label{fig:SOD2D}
\end{figure}

We have used NVIDA Nsight Systems to trace an execution of SOD2D with 2 and 4 nodes, for 11 steps,  and translated it with nsys2prv~\cite{nsys2prv,nsys2prv_sw} to visualize it in Paraver. 
We reduced the number of steps for the detailed tracing because using the same number of iterations as in the TALP profiling would generate an unmanageable volume of data.
Figure~\ref{fig:SOD2D}, shows the CUDA kernel for the whole execution, where we observe three distinct regions of code:  (a) \texttt{initialization}, in a green box and at the beginning, when the data is copied from Host-to-Device (b) \texttt{one time step} (marked with a purple box in the image) executed after a large gap where the CPU is busy initializing everything and there is no activity in the GPU, and lastly (c) \texttt{11 Timesteps}, shown in a blue rectangular shape in the top image.

To better grasp what TALP observed with 1001 steps, Figure~\ref{fig:SOD2D}, bottom image, shows the CUDA kernel executions of 1 step  using 8 devices, each row corresponds to one CUDA device. In this trace we see that during the timestep execution the GPUs are active nearly all the time.

As SOD2D does all of its computation in the GPUs, the low and steady Device Offload Efficiency and a high Orchestration Efficiency is also understandable.

\begin{figure}[htb!]
    \centering
    \includegraphics[width=0.9\columnwidth]{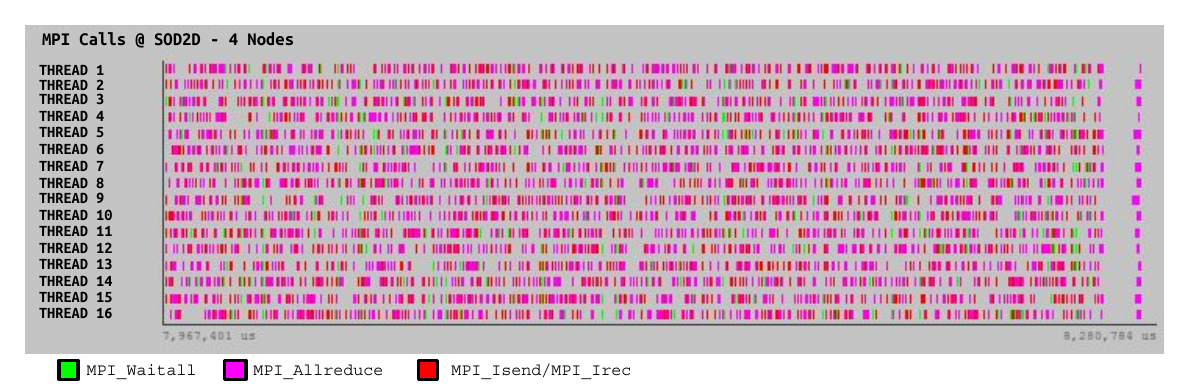}
    \caption{SOD2D trace showing MPI calls on 4 nodes, 16 processes.}
    \label{fig:sod2d_MPI_calls}
\end{figure}

Figure~\ref{fig:sod2d_MPI_calls} shows the MPI calls executed across 4 nodes and 16 GPUs. The number of calls is substantial and can be interpreted as an obstacle that prevents threads from spending more time on computation or on offloading work to the GPUs. Consequently, the drop in MPI Communication Efficiency on the host side, as observed in TALP output, is expected.

\subsubsection{FALL3D}
.\\

Fall3D is an Eulerian model for atmospheric passive transport and deposition, formulated on the advection--diffusion--sedimentation (ADS) equation. The model has been extended with new capabilities, including support for ensemble forecasts and the simulation of multiple atmospheric species (e.g., volcanic ash, gases, mineral dust, and radionuclides). FALL3D is implemented in Fortran, with parallelization achieved through MPI and CUDA.

\begin{table}[!htbp]
    \centering
    \caption{TALP Output for FALL3D from 1 to 8 nodes}
    \label{tab:fall3d}
    \begin{tabular}{p{.5cm}p{4cm}GGGG}
        \toprule
        & \multicolumn{1}{l}{Metrics} & \multicolumn{4}{c}{ Nodes} \\
        \cmidrule(lr{0.5em}){3-6}
        & 
        & \multicolumn{1}{c}{1}
        & \multicolumn{1}{c}{2}
        & \multicolumn{1}{c}{4}
        & \multicolumn{1}{c}{8}    \\
        
        \midrule
         \multirow{7}{*}{\rotatebox{90}{Host}}
         & \texttt{Parallel Efficiency}             & 0.26 &  0.16 &  0.10 &  0.07 \\
        & \texttt{├─ MPI Parallel Eff.}       & 0.44 &  0.27 &  0.16 &  0.11 \\
        & \texttt{│\ \ ├─ Comm. Eff.} & 0.85 & 0.83 & 0.82 & 0.89 \\
        & \texttt{│\ \ └─ Load Balance}             & 0.52 &  0.32 &  0.20 &  0.12 \\
        & \texttt{└─ Device Offload Eff.}     & 0.59 & 0.61 & 0.63 & 0.64 \\
         \midrule

         \multirow{4}{*}{\rotatebox{90}{Device}}
        & \texttt{Parallel Efficiency}              & 0.14 & 0.08 & 0.04 & 0.03 \\
        & \texttt{├─ Load Balance}                  & 0.98 & 0.97 & 0.96 & 0.96 \\
        & \texttt{├─ Communication Eff.}      & 0.78 & 0.77 & 0.76 & 0.75 \\
        & \texttt{└─ Orchestration Eff.}      & 0.19 & 0.11 & 0.06 & 0.04 \\
        \bottomrule
    \end{tabular}
\end{table}

In Table~\ref{tab:fall3d} we can see the efficiency metrics obtained with TALP of FALL3D runs between 1 and 8 nodes. 
The TALP output highlights several performance bottlenecks in FALL3D. The MPI Parallel Efficiency is notably low, primarily due to load imbalance. 
At the host level we also observe a low Device Offload Efficiency, meaning that the CPUs are not used to do useful work while the GPUs are executing kernels.

\begin{figure}[bht!]
    \centering
    \includegraphics[width=0.9\columnwidth]{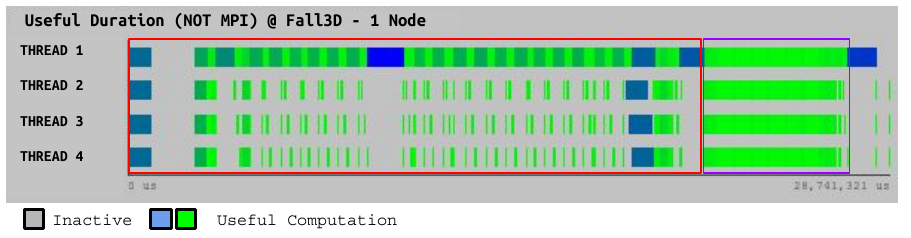}
    \includegraphics[width=0.9\columnwidth]{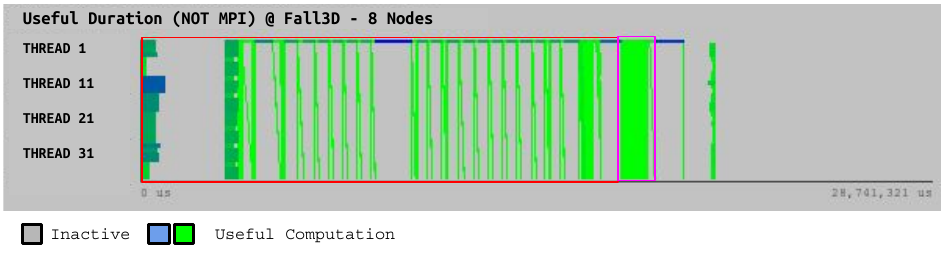}
     \caption{FALL3D traces showing useful duration events. Top: 1 node, 4 GPUs. Bottom: 8 nodes, 32 GPUs.}
    \label{fig:FALL3D}
\end{figure}

At the device level we observe a low Orchestration Efficiency, meaning that not enough work is being offloaded to the GPUs to keep them busy doing useful work. Moreover
as the application scales on more nodes, Orchestration Efficiency decreases, which can be attributed to a decreasing MPI Parallel Efficiency. The host is not able to provide work to the GPUs due to the MPI inefficiencies.
Conversely, Device Offload Efficiency increases, indicating that CPUs spend proportionally less time in the CUDA runtime.

\begin{figure}[htb!]
    \centering
    \includegraphics[width=0.9\columnwidth]{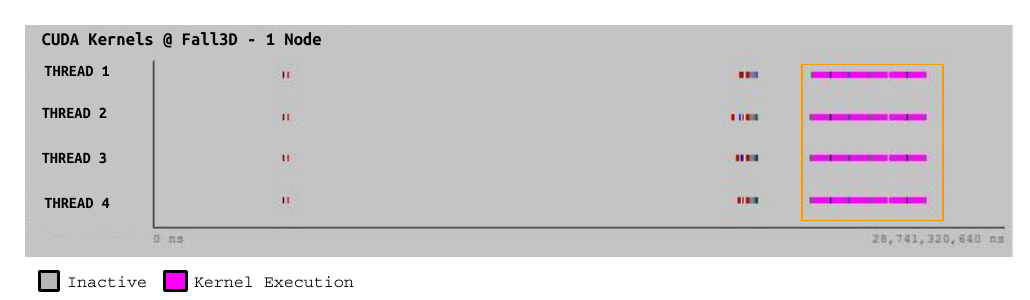}
    \includegraphics[width=0.9\columnwidth]{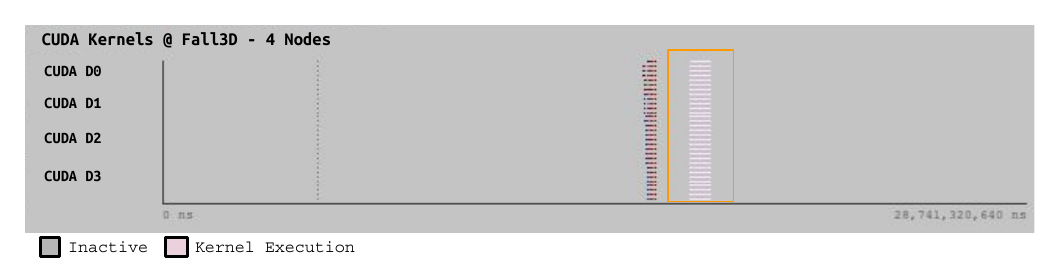}
    \caption{FALL3D traces showing CUDA Kernel execution. Top: 1 node, 4 GPUs. Bottom: 8 nodes, 32 GPUs.}
    \label{fig:FALL3D-CUDA}
\end{figure}

 For the verification of TALP outputs, we have traced FALL3D, for 1 and 8 nodes as seen in the two windows in figure~\ref{fig:FALL3D} and figure~\ref{fig:FALL3D-CUDA}. Both of the windows in both figures, show the same time scale. 
 In figure~\ref{fig:FALL3D}, each row corresponds to one MPI process (we omit the GPUs for clarity), red boxes in both traces mark the initialization and purple boxes mark the iterative region of the application. 
 The load imbalance in host is visible in both windows, where MPI process, rank 0 (labeled thread 1) spends most of the duration of initialization in distributing the workload among the other MPI processes. As the number of resources increases, the load balance becomes more visible in Fall3D.  
 
 It is visible from figure~\ref{fig:FALL3D-CUDA}, that the CUDA Kernel executions, seen in a yellow boxes in both images on top and bottom, on devices scale from 1 to 8 nodes. Therefore, the increase in Device Offload Efficiency is justified as the number of GPUs increase and  consequently, the kernel execution time decreases. As the kernel executions scale with increasing the number of the nodes, GPUs spend less time in computation and therefore, the Orchestration Efficiency decreases. 
 Finally, due to the long initialization phase and that GPUs are not used during the initialization, GPUs spend more of their time in inactive state while scaling the application. This is identifiable in low numbers seen in Orchestration Efficiency in TALP output.

\subsubsection{XSHELLS}
.\\

Xshells is a high performance simulation code for the rotating Navier-Stokes equation in spherical shells, optionally coupled to the induction and temperature equation. XSHELLS is written in C\texttt{++} and uses MPI for distributed memory parallelism and CUDA for GPU offloading.

\begin{table}[!htbp]
    \centering
    \caption{TALP Output for XSHELLS from 1 to 8 nodes}
    \label{tab:xshells}
    \begin{tabular}{p{.5cm}p{4cm}GGGG}
        \toprule
        & \multicolumn{1}{l}{Metrics} & \multicolumn{4}{c}{ Nodes} \\
        \cmidrule(lr{0.5em}){3-6}
        & 
        & \multicolumn{1}{c}{1}
        & \multicolumn{1}{c}{2}
        & \multicolumn{1}{c}{4}
        & \multicolumn{1}{c}{8}    \\
        
        \midrule
         \multirow{7}{*}{\rotatebox{90}{Host}}
& \texttt{Parallel Efficiency}           & 0.36  & 0.29  & 0.26 & 0.15 \\
        & \texttt{├─ MPI Parallel Eff.}  & 0.90 & 0.64 & 0.51 & 0.25 \\
        & \texttt{│\ \ ├─ Comm. Eff.}    & 0.91 & 0.66 & 0.52 & 0.27 \\
        & \texttt{│\ \ └─ Load Balance}  & 0.98 & 0.97 & 0.98 & 0.93 \\
        & \texttt{└─ Device Offload eff.} & 0.40 & 0.45 & 0.51 & 0.60 \\
         \midrule

         \multirow{4}{*}{\rotatebox{90}{Device}}
         & \texttt{Parallel Efficiency}  & 0.52 & 0.35 & 0.24 & 0.10 \\
        & \texttt{├─ Load Balance}       & 1.00 & 1.00 & 1.00 & 1.00 \\
        & \texttt{├─ Communication Eff.} & 0.98 & 0.97 & 0.96 & 0.94 \\
        & \texttt{└─ Orchestration Eff.} & 0.54 & 0.36 & 0.25 & 0.10 \\
        \bottomrule
    \end{tabular}
\end{table}

We can find the output of TALP when running XSHELLS on 1 to 8 nodes, as shown in Table~\ref{tab:xshells}. The TALP output shows that the computational load is well balanced across CPUs and GPUs as Load Balance values from both the host and the device are very high. 

On the other hand, the Host Communication Efficiency decreases abruptly when scaling from 1 to 8 nodes. In the host side we also observe that using one node the Device Offload Efficiency metric is around 40\% telling us that 60\% of the time the host is doing useful work as opposed to the 40\% of the time waiting for the GPU, we also see that this metric increases as we increase the number of nodes. This means that the amount of work done by the CPUs increase as we scale the number of nodes

In the device metrics, we observe good Load Balance and Communication Efficiency, being Orchestration Efficiency the only metric showing low values.

The Orchestration Efficiency starts at an already low value of 54\% for a single node and drops to 10\% with 8 nodes. This decrease in the overall device efficiency is primarily attributed to low host-side metrics, specifically the Device Offload Efficiency and MPI Communication Efficiency. The host is likely busy with MPI communication and CPU useful work, therefore unable to offload enough work to keep the GPUs more utilized.

\begin{figure}[bht!]
    \centering
    \includegraphics[width=0.9\columnwidth]{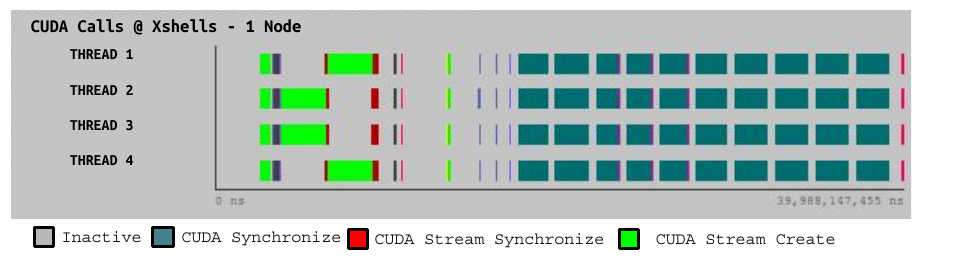}
    \includegraphics[width=0.9\columnwidth]{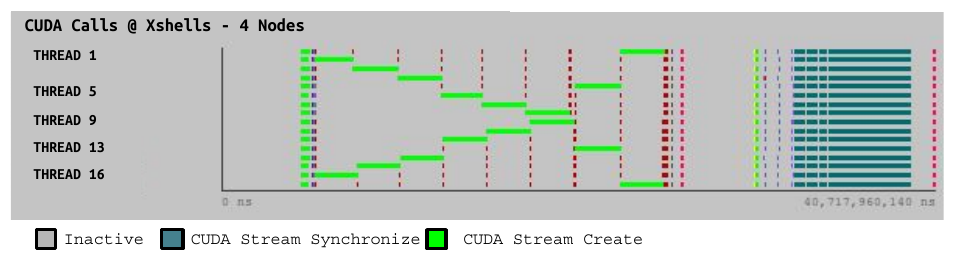}
    \caption{XSHELLS traces showing CUDA calls. Top: 1 node, 4 GPUs. Bottom: 4 nodes, 16 GPUs.}
    \label{fig:XSHELL}
\end{figure}

\begin{figure}[htb!]
    \centering
    \includegraphics[width=0.9\columnwidth]{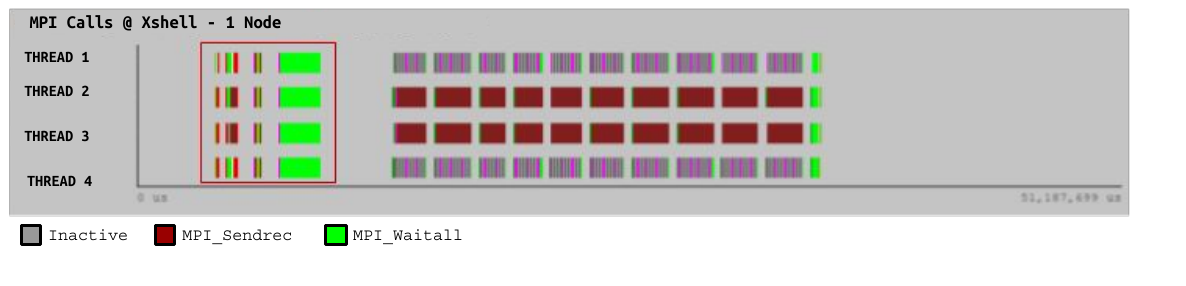}
    \includegraphics[width=0.9\columnwidth]{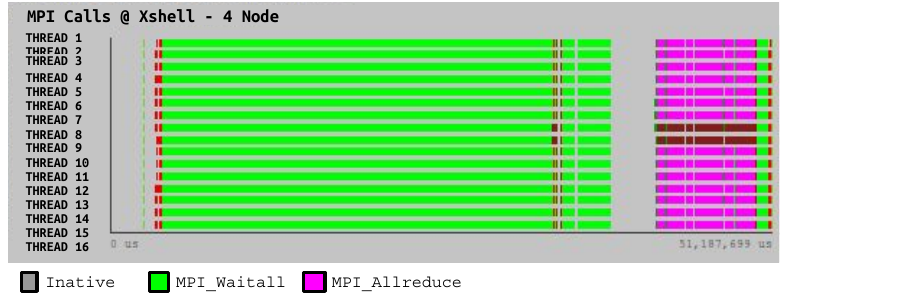}
    \caption{XSHELLS traces showing MPI calls. Top: 1 node, 4 GPUs. Bottom: 4 nodes, 16 GPUs.}
    \label{fig:XSHELL_MPI_Call_1N_4N}
\end{figure}

In Figure~\ref{fig:XSHELL} we show two traces of XSHELLS, in each one we show only the MPI ranks when doing CUDA calls for 1 node at the top and 4 nodes at the bottom. Both windows have the same execution time and scale. 

The top trace of Figure~\ref{fig:XSHELL} for a single node illustrates that the host is only offloading work to the GPUs (turquoise color) for approximately half of the total timeline. This visually corresponds to the measured Orchestration Efficiency of 54\%  resported on a single node. 

In the bottom trace that corresponds to the 8 node execution, we see that the initialization phase does not scale well and therefore takes longer in 4 nodes than in 1, while the iterative phase launching kernels scale well and its time is reduced when increasing the number of nodes. In the case of the 4 nodes trace we see that the iterative phase takes a quarter of the time and corresponds to the 25\% Orchestration Efficiency observed in 4 nodes.

In Figure~\ref{fig:XSHELL_MPI_Call_1N_4N} we present two traces that show the MPI calls done by the host, each row corresponds to an MPI rank, and the color represents the MPI call that is being executed. The top trace corresponds to 1 node (4 MPI ranks) execution and the bottom trace to 4 nodes (16 MPI ranks execution). We have marked with a red rectangle the initialization phase.
 
We observe that there is an MPI intensive phase in the initialization that is not scaling, and this is identified by the decrease in the MPI Communication Efficiency.

\section{Conclusions and Future Work}
\label{sec:conclusions}

Accelerators such as GPUs play an increasingly important role in High Performance Computing (HPC). However, using them efficiently in large-scale HPC codes remains a challenging task, making performance measurement and analysis crucial. Consequently, the introduction of new GPU-specific metrics is essential.

In this work, we propose novel metrics that extend the existing POP efficiency framework to separately capture host and device behavior in heterogeneous architectures. By separating efficiency metrics, performance analysis in accelerated applications becomes simpler and more effective. Distinguishing between host and device metrics allows a clearer understanding of execution patterns and facilitates analysis aligned with developers' intent for the targeted hardware.

The proposed metrics are vendor-independent and applicable across accelerator platforms such as AMD and NVIDIA GPUs, ensuring their relevance to modern HPC clusters and supercomputers. Specifically, we introduced one host-side metric, Device Offload Efficiency, and three device-side metrics: Load Balance, Communication Efficiency, and Orchestration Efficiency.

We implemented these metrics within the TALP framework, enabling HPC users to obtain efficiency measurements for accelerated platforms in a transparent and lightweight manner. The current implementation supports codes running on NVIDIA GPUs with CUDA or OpenACC. As future work, we plan to implement the second part of the graph, or Computational Efficiency and extend the implementation to AMD GPUs to broaden applicability across accelerators.

Finally, we evaluated the proposed metrics and their TALP integration using both synthetic benchmarks and three scientific production codes. Synthetic benchmarks allowed us to model controlled workload and distribution patterns, while production codes demonstrated real-world applicability. In both cases, we compared the metrics against detailed execution traces, validating that the proposed metrics provide accurate and actionable insights into performance behavior.

\subsubsection*{\ackname}
This work has received funding from the European High Performance Computing Joint Undertaking (JU) and Spain, Italy, Iceland, Germany, Norway, France, Finland and Croatia under grant agreement No 101093038, ChEESE-2P, project PCI2022-134980-2 funded by MCIN/AEI/10.13039/501100011033 and by the European Union NextGenerationEU/PRTR. 

This work has received funding from the European High Performance Computing Joint Undertaking (JU) and Spain, Germany, France, Portugal, and the Czech Republic under grant agreement No 101143931, POP3, project PCI2024-153419 funded by MICIU/AEI.

%
\bibliographystyle{splncs04}
\bibliography{bibliography}

\end{document}